\documentclass{emulateapj}
\usepackage{natbib}

\usepackage{graphicx}
\usepackage[space]{grffile}
\usepackage{latexsym}
\usepackage{amsfonts,amsmath,amssymb}
\usepackage{url}
\usepackage[utf8]{inputenc}
\usepackage{fancyref}
\usepackage{hyperref}
\hypersetup{colorlinks=false,pdfborder={0 0 0},}
\usepackage{textcomp}
\usepackage{longtable}

\usepackage{enumitem}

\def\hkpc{h^{-1}\mathrm{kpc}}
\def\hMpc{h^{-1}\mathrm{Mpc}}

\def\gt{>}

\shorttitle{Matching analytical models of cosmic reionization with numerical simulations}
\shortauthors{Alexander A. Kaurov}

\begin{document}


\title{On improving analytical models of cosmic reionization \\ for matching  numerical simulation}

\author{Alexander A.\ Kaurov\altaffilmark{1}}
\altaffiltext{1}{Department of Astronomy \& Astrophysics, The
  University of Chicago, Chicago, IL 60637 USA; kaurov@uchicago.edu}


%\altaffiltext{1}{Visiting Astronomer, Cerro Tololo Inter-American Observatory.
%CTIO is operated by AURA, Inc.\ under contract to the National Science
%\altaffiltext{2}{Society of Fellows, Harvard University.}
%\altaffiltext{3}{present address: Center for Astrophysics,
%    60 Garden Street, Cambridge, MA 02138}
%\altaffiltext{4}{Visiting Programmer, Space Telescope Science Institute}
%\altaffiltext{5}{Patron, Alonso's Bar and Grill}


\begin{abstract}
The methods for studying the epoch of cosmic reionization vary from full radiative transfer simulations to purely analytical models. While numerical approaches are computationally expensive and are not suitable for generating many mock catalogs, analytical methods are based on assumptions and approximations. We explore the interconnection between both methods. First, we ask how the analytical framework of excursion set formalism can be used for statistical analysis of numerical simulations and visual representation of the morphology of ionization fronts. Second, we explore the methods of training the analytical model on a given numerical simulation. We present a new code which emerged from this study. Its main application is to match the analytical model with a numerical simulation. Then, it allows one to generate mock reionization catalogs with volumes exceeding the original simulation quickly and computationally inexpensively, meanwhile reproducing large scale statistical properties. These mock catalogs are particularly useful for CMB polarization and 21cm experiments, where large volumes are required to simulate the observed signal. 
\end{abstract}

\bibliographystyle{apj}

\section{Introduction}

The nature of the epoch of cosmic reionization involves a large dynamic range of scales and various physical processes. An accurate treatment of this epoch requires detailed bookkeeping of the photon budget. Therefore, good understanding of astrophysics as well as large scale structure formation and cosmology is a necessity. Indeed, after being emitted by a star, a photon travels through partially neutral interstellar medium (ISM) and then through intergalactic medium (IGM) populated with Lyman limit systems (LLS) before it reaches the ionization front. The diversity of these environments makes study of reionization challenging.

The past two decades have seen rapid development of analytical \citep{miralda2000reionization,Furlanetto_2004,Kuhlen2012} and numerical \citep{Iliev2006,McQuinn2007,Zahn2007, Croft2008,Lee2008,Shin2008,Trac2008, Iliev2009,Aubert2010,Friedrich2011, Ahn2012,Shapiro2012,Hutter2014, Iliev2014,So2014,Norman2015} methods of studying the epoch of cosmic reionization. Nevertheless, both of them struggle from various limitations. The growth of computer capabilities gave a great boost to the numerical methods \citep{Trac2011}. However, still, simulations are limited by the dynamic range, i.e. either smaller boxes with high resolution or large boxes with low resolution. Therefore, either the correlations on large scales are neglected or the crude approximations are adopted for sub-grid models. 

On the other hand, the analytical models are computationally inexpensive and potentially can probe the volumes of any sizes. Also, the framework of these models provides a descriptive view on reionization, and therefore develops intuition about the significance of each physical process. However, those assumptions, which make the framework so efficient, also cause some limitations. Firstly, it limits physical processes which can be included in the model. Secondly, the approximations cause unpredictable accuracy. In particular, one cannot invest more computational power to achieve higher accuracy.

These limitations are reflected in the capabilities of these models to be compared with observations. For instance, in order to match the galaxy luminosity function (star formation history) and escape fractions one has no other choice but to run a detailed high resolution simulation; while for the large scale morphology studies (21cm tomography) large number of realizations in cosmological volumes are needed, which can be done only with approximate semi-numerical models.
 
In this study we attempt to take the best from both methods. First, in \S\ref{sec:analitic_for_numerical} we approach the numerical simulation as if it was a product of semi-analytical computation. As a result, we end up with a new kind of statistics based on excursion set formalism, which describes the ionization field on large and small scales. This tool allows one to visualize the dependence of ionization history on the underlying density perturbations in an unusual way.

Then, in \S\ref{sec:tuning} we test how descriptive is the phase space statistics calculated in \S\ref{sec:analitic_for_numerical}. If it is informative enough, then it will be sufficient for reproducing the reionization history on a different set of initial conditions. In other words, we perform a training (or tuning) of an analytical model on a given numerical simulation and discuss how well it can be done. In contrast to \cite{Battaglia2013} where a similar goal was set, we use the excursion set formalism \citep{Press1974} instead of bias approach. The performance of the excursion set analytical models versus numerical simulation has been carried out in \citet{Zahn2011}, where the authors put the same physics (the efficiency of ionizing photon production) into numerical and analytical models. We adopt a different approach, and try to find \textit{any} analytical model that would correspond to a given numerical simulation.

The results from this study allow us to make conclusions about capabilities of the analytical approach, i.e. what physical processes they generally fail to mimic and for which tasks it can be used for. Also, we show how the analytical framework can play a complimentary role in numerical calculations by providing a tool for visualization of spatially complicated reionization history.

Additionally, we present a semi-numerical code based on excursion set formalism which emerged from this study. Its main application is training of an analytical model on a given numerical simulation. Then, this trained model can be used for rapid generation of mock catalogs for 21cm experiments.

We start by discussing the motivation in \S\ref{sec:discussion} and giving a brief overview in \S\ref{sec:preliminaries} of the framework of the two main analytical approaches we consider (\S\ref{subsec:analyticmodels}) and the numerical simulation that we use as our fiducial model (\S\ref{subsec:numerical_simulation}). Those are followed by \S\ref{sec:analitic_for_numerical} and \S\ref{sec:tuning} where we present and discuss our method. In \S\ref{sec:conclusions} we conclude.

%----------------------
\section{Motivation}
\label{sec:discussion}
%----------------------

The value of any theoretical model is its capability to be experimentally tested. The new observational facilities are able to probe the most interesting for reionization range of redshifts $5<z<20$. Here we discuss how those can be estimated in the theoretical models. First, we list those observations which can be compared versus existing numerical simulations:  

\begin{itemize}[leftmargin=*]
  \item The integrated optical depth of Thomson scattering on free electrons to the cosmic microwave background provides a glimpse to the history of ionization. The measured value by the Wilkinson Microwave Anisotropy Probe argues in favor of the scenario when reionization began at $z\gt15$ \citep{Hinshaw_2013,Bromm_2011,Dunlop_2012} with measured optical depth: $\tau=0.081\pm0.012$. The measured optical depth by \citet{PlanckCollaboration2015} is lower: $\tau = 0.066\pm0.016$. This observable can be matched with global ionization history; hence, even models of homogeneous reionization can simulate this quantity.

  \item The end of reionization is probed by the observation of the Gunn-Peterson effect in spectrum of high redshift QSOs \citep{Fan_2006,Bolton_2011} and gamma gay bursts \citep{Chornock_2013}, as well as the Lyman-$\alpha$ emission from high redshift galaxies \citep{Stark_2010, Pentericci_2011,Pentericci_2014,Schenker_2011,Schenker_2014,Treu_2013,Tilvi_2014}.
  
  \item Another observable is luminosity function and its derivative star formation rate at high redshift. Current observations by Hubble Space Telescope Ultra Deep allow to probe the galaxies up to redshift 10. The modern simulations of reionization including the one we use \citep{Gnedin2014a} are capable of reproducing these luminosity functions.

\end{itemize}

In contrast to the observations listed above, these cannot be numerically simulated with the same precision due to much larger volumes:

\begin{itemize}[leftmargin=*]
  \item The polarization of CMB caused by reionization allows to put constraints on its duration and redshift \citep{Zahn2012, Benson2014}.
    
  \item The most intriguing observation is 21 cm line, which will allow to map hydrogen at high redshifts. Those include SKA\footnote{http://www.skatelescope.org/}, LOFAR\footnote{http://www.lofar.org/}, MWA \cite{Tingay2013}, PAPER \citep{Parsons2010} and HERA\footnote{http://reionization.org/}.
\end{itemize}

The data from these types of experiments require Monte Carlo simulations for proper analysis. However, current computational capabilities do not allow to run multiple simulations in large (1 Gpc) cosmological boxes with fine radiation transfer and galaxy formation. Therefore, semi-numerical models are adopted for these tasks.

%----------------------
\section{Preliminaries} 
\label{sec:preliminaries}
%----------------------

%----------------------
\subsection{Analytic models}
\label{subsec:analyticmodels}
%----------------------

We leave beyond the scope of this paper the analytical models of homogeneous reionization (i.e. \cite{Kuhlen2012}) and focus on the two of the most popular inhomogeneous models. Those are \cite{miralda2000reionization} (\textbf{MHR00} hereafter) and \cite{Furlanetto_2004} (\textbf{FZH04} hereafter) and their derivatives.

Both models are based on the underlying overdensity field $\delta$. The key concept is the density scale, i.e. the density averaged over some scale $R$. The density $\rho(R)$ can be defined as:
\begin{equation}
\rho(R) = \int_{V_R} \delta \mathrm{d}V / V_R,
\end{equation}
where $V_R$ is the volume of a sphere of a radius $R$, centered at the point of interest. However, in practice smoothing is performed through filtering in the Fourier space. 

If the Gaussian density field is considered, then another useful quantity emerges from the Fourier definition -- the variance of density at a given scale:
\begin{equation}
\label{eq:sigmaR}
\sigma(R)^2 = \left\langle \delta(R)^2 \right\rangle = 
\int_0^\infty P(k) \tilde{W}^2 (kR)\;\mathrm{d}^3k,
\end{equation}
where $P(k)$ is the power spectrum and $\tilde{W}_R$ is the Fourier transform of a spherical filter:
\begin{equation}
W_R(r) = \Theta(1-r/R).
\end{equation}
%Further details about the filtering method used in this paper are discussed in Appendix \ref{app:fourier_filter}.

We do not specify what the scalar field, $\delta$, is that we use in this method. It can be the baryon or dark matter density, the initial Gaussian field or evolved non-linear field. Also, it can be the halo density field or star formation density.

In case of the Gaussian field the distribution of $\rho(R_0)$ in the universe is Normal by definition for any $R_0$, and $\sigma(R_0)^2$ calculated with the Equation \ref{eq:sigmaR} is its variance. However, for any other non-Gaussian scalar field, the distribution can diverge from Normal. Nevertheless, in order to have the plots easily readable and consistent with each other we normalize the $\rho(R_0)$ to zero mean and unit variance. Even though this operation erases the shape of distribution function, it essentially affects only visual representation of the figures. Thus, we label those normalized values as ``in units of $\sigma^2$" in all the figures.

It is common in the literature to use other variables instead of scale $R$. Among those are the mass, $m$, of a sphere with the radius $R$, or variance, $\sigma^2$, at scale $R$. All three values can be derived from each other. In this paper we use only $R$ since it is most appropriate for our purposes.

%----------------------
\subsubsection{MHR00 model}
%----------------------

The MHR00 model was built to describe the later stages of reionization when only the small patches of hydrogen are left neutral. It describes a process of the ionization background burning into the dense neutral regions, where the recombination and photoionization rates are comparable. The characteristic overdensities of the regions where this effect takes place were studied in \cite{2014arXiv1412.5607K} with the same simulation. The result shows that inside the ionized regions the local density is a proxy for the local ionization fraction.

This model uses one specific scale, $R_\mathrm{MHR}$, and determines the ionization fraction of a cell based on its local density $\rho(R_\mathrm{MHR})$ only. This relatively simple model is capable of describing Lyman-$\alpha$ forest statistics.

The model is based on the assumption of quasi-static ionization equilibrium, i.e. when the local recombination rate is close to the ionization rate:
\begin{equation}
R_u x_\mathrm{ion}^2 (1+\delta)^2 = \Gamma(1+\delta)(1-x_\mathrm{ion}),
\end{equation}
where $\delta$ and $x_\mathrm{ion}$ are local gas overdensity and ionizaed fraction, $\Gamma$ is ionizing background, and $R_u$ is the ratio of the recombination rate for a homogeneous universe to the Hubble constant, which is order of unity at $z\approx 6$ (see MHR00, Equation 1). Solving this equation for given ionization background and $x_\mathrm{ion}=0.5$ gives the density threshold above which the gas is neutral.

Even though this model describes the inhomogeneous reionization (some regions are ionized later than other) it still assumes the homogeneous ionization background throughout the universe. Thus, the main limitation of the model is that it does not incorporate any correlation from outside of $R_\mathrm{MHR}$ radius and therefore does not depend on the Large-Scale Structure.

%----------------------
\subsubsection{FZH04 model}
\label{subsec:FZH04}
%----------------------

In contrast to the MHR00, the FZH04 model was developed to describe the morphology of the reionization on the large scales. In order to reproduce the large scale inhomogeneity it has to take into account overdensity not only at a single scale but on a range of scales.

In the FZH04 model we assign to each point at position $\textbf{r}$ a one-dimensional function, trajectory $T(R)$, which corresponds to the density of the point defined at different scales $R$:
\begin{equation}
T_\textbf{r}(R) = \dfrac{ \int \delta(\textbf{x}) W_R(|\textbf{r}-\textbf{x}|)\mathrm{d}\textbf{x}}{ \sigma(R)^2 }.
\end{equation}

The physical motivation is that the function $T(R)$ can provide an estimate for the fraction of matter collapsed into halos \citep{Press1974}. Then, assuming a rate of ionizing photons production in halos, one can derive the total number of ionizing photons in the spherical regions centered at the point of interest. If for any of those regions the number of ionizing photons exceeds the number of hydrogen atoms in the same region, the point is considered ionized.

Various improvements can be made for this model, which allow to include more sophisticated physics \citep{Furlanetto2005,Furlanetto2006,Alvarez2007,Mesinger2007,Mesinger2011,Zahn2011,Alvarez2012,Battaglia2013,Kaurov2013,Zhou2013,Kaurov2014,Sobacchi2014}. However, the main principle remains the same. All physics are combined into the main parameter of the model --  the \textit{barrier} function, $B(R,\,z)$, which is the function of scale and redshift. The model predicts that the point is ionized at redshift $z$ if its trajectory $T(R)$ intersects with $B(R,\,z)$. 

This model gained popularity because it is physically justified, based on excursion set framework which is widely used in structure formation theory, and is computationally efficient. It became a starting point for the development of semi-numerical codes like 21cmFAST \citep{Mesinger2011}.

The weak part of this family of models is the transition from physical equations to the shape of $B(R,\,z)$. At this step a lot of assumptions are made and it is hard to trace how they affect the accuracy of the model. The barrier effectively becomes degenerate because of the large amount of physical effects contributing to its shape.

%---------------------------------
\subsection{Numerical simulation}
\label{subsec:numerical_simulation}
%---------------------------------

For this study we use simulations from the Cosmic Reionization On Computers (CROC) project \citep{Gnedin2014a, Gnedin2014b} as a reference numerical simulation. These simulations include a large variety of physical processes which affect the morphology of reionization in different ways on various scales. For instance, the non-spherically-symmetric escape fraction may affect the shape of small bubbles at earliest stages of reionization, while the bias of galaxy distribution regulates the morphology of ionization fronts on large scales.

The simulation is tuned to match the available observational constraints. Those include the galaxy luminosity functions and full distribution function of Gunn-Peterson absorption in the spectra of the high redshift quasars. However, in this study we do not need precise tuning of a simulation. We only need a complexity of the ionization bubbles and the shapes of neutral patches, in order to test our methods.

With the use of Adaptive Mesh Refinement algorithm, CROC simulations achieve spatial resolution of $125\mathrm{pc}$ in simulation volumes of up to $40\hMpc$, which allows to consider IGM and filaments to be well resolved. 

In this paper we use a $40\hMpc$ realization (run B40.sf1.uv2.bw10.A from \citet{Gnedin2014a}) as our fiducial set and two other $40\hMpc$ realizations (runs B40.sf1.uv2.bw10.B,C from \citet{Gnedin2014a}) for comparison. The scalar fields which we extract from simulation are: the baryon and dark matter density fields, the ionization fractions of Hydrogen.

%----------------------
\section{The framework of analytical models for analysis of numerical results}
\label{sec:analitic_for_numerical}
%----------------------

In this section we approach a realization of the numerical simulation with the framework of the analytical model.

\begin{figure*}
\begin{center}
\includegraphics[width=0.95\columnwidth]{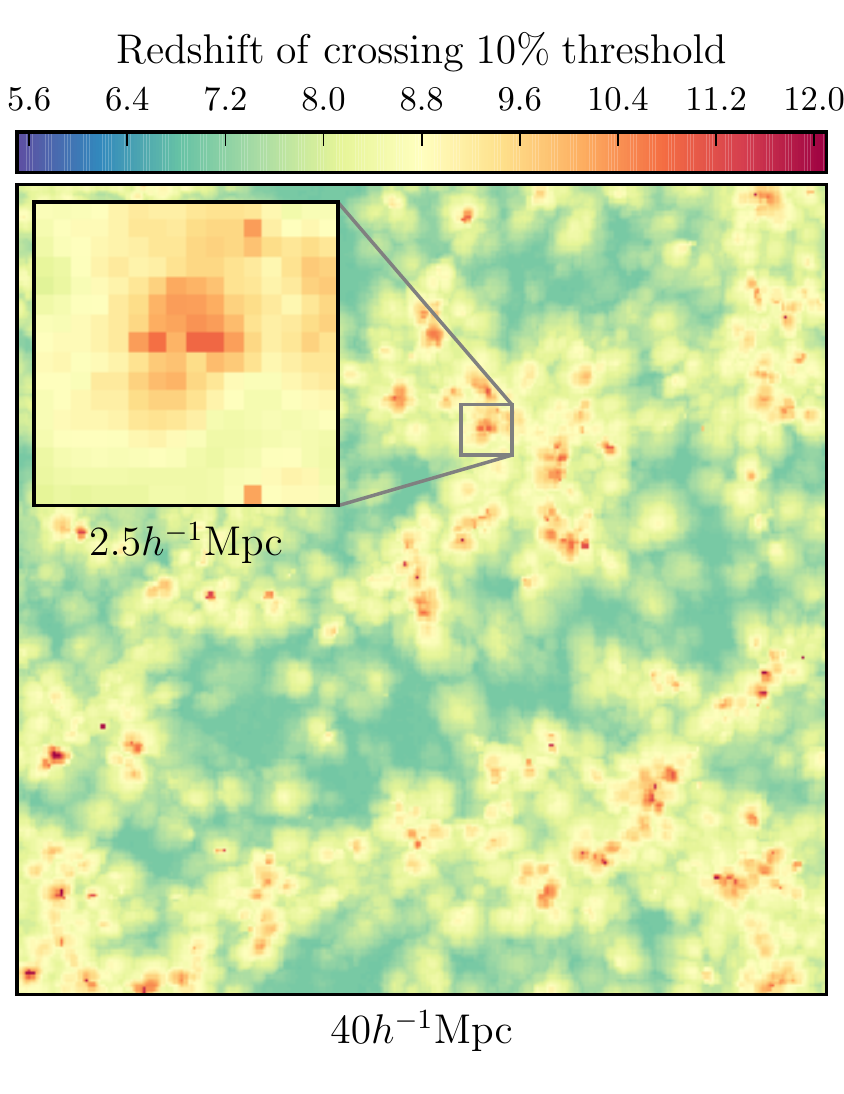}
\includegraphics[width=0.95\columnwidth]{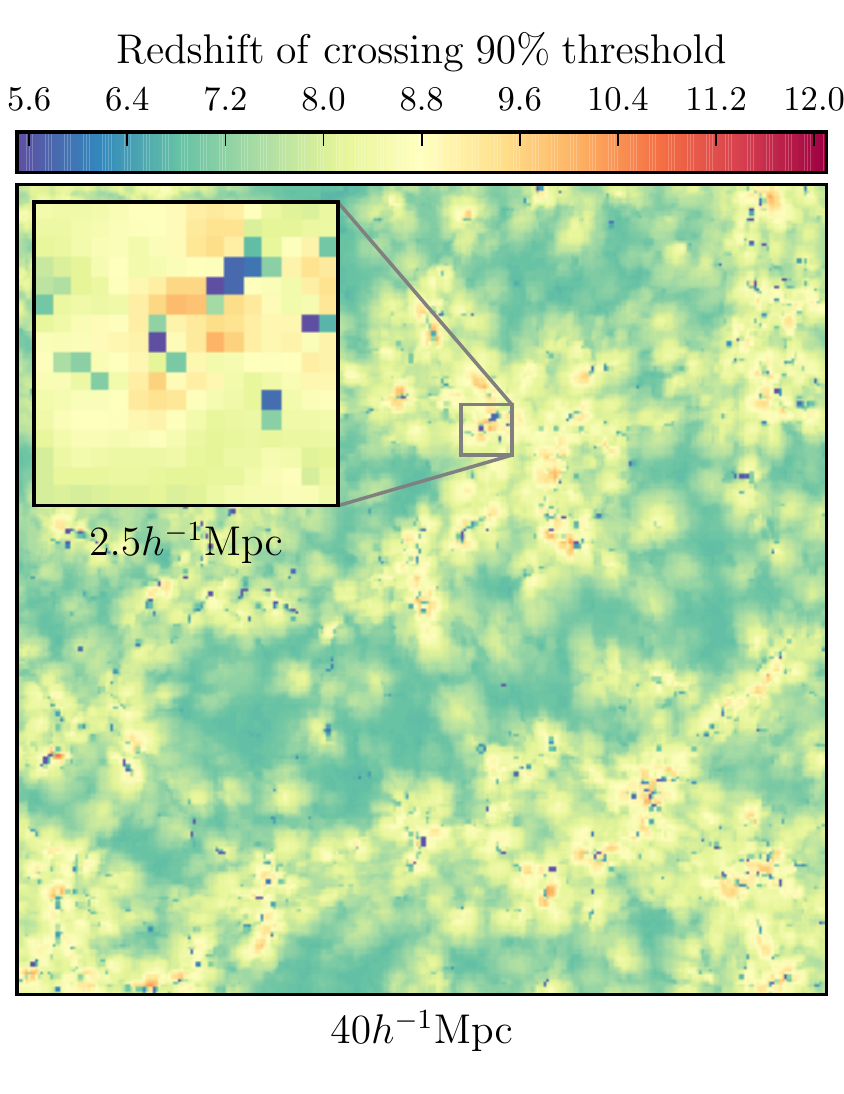}
\caption{\label{fig:slice-reionization} The moment of reaching 10\% (left panel) and 90\% (right panel) ionization level in a slice from $40\hMpc$ simulation.%
}
\end{center}
\end{figure*}

\begin{figure*}
\begin{center}
\includegraphics[width=0.95\columnwidth]{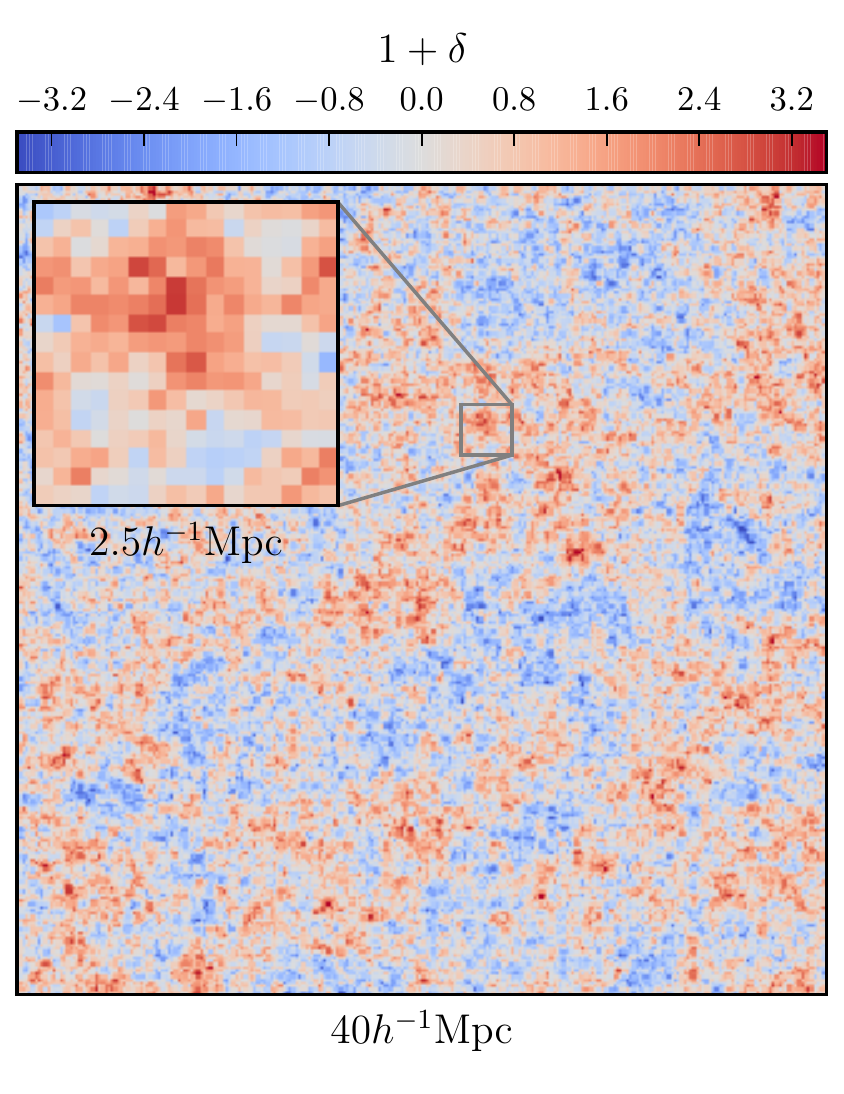}
\includegraphics[width=0.95\columnwidth]{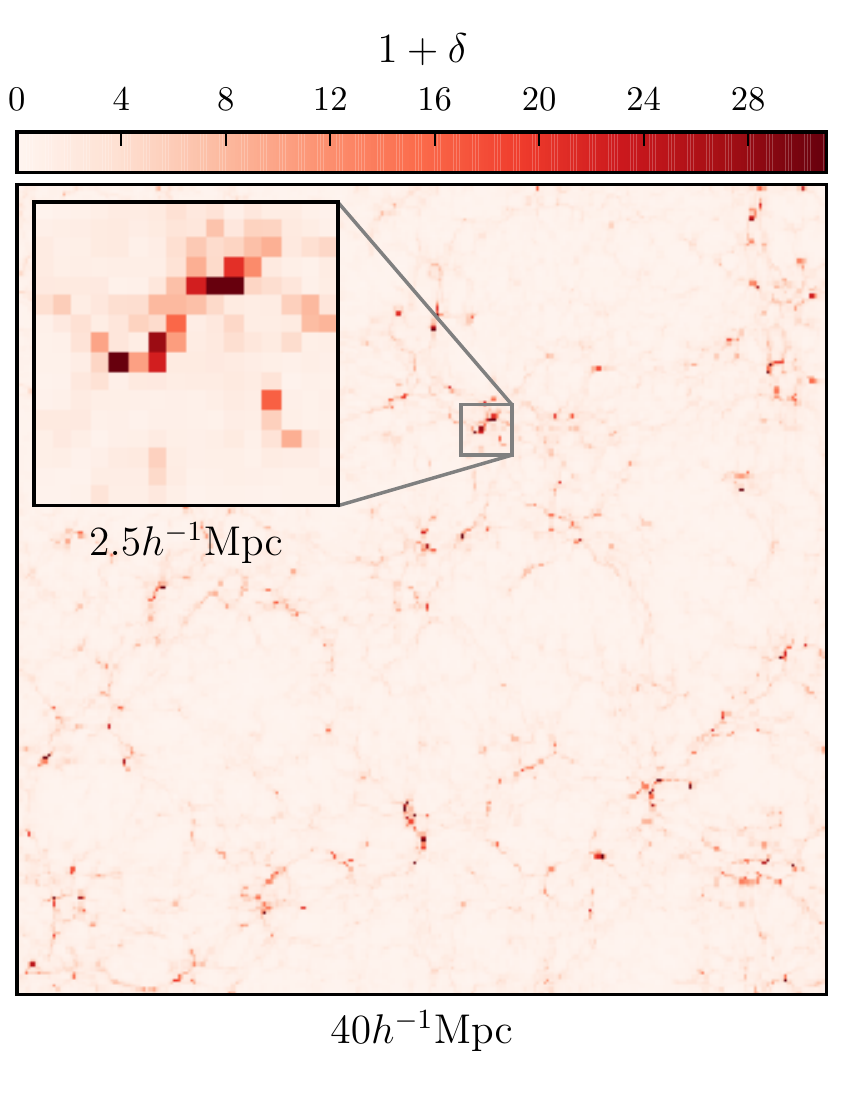}
\caption{\label{fig:slice-delta} The overdensity of initial conditions (left panel) and evolved baryon density field  at redshift 6 (right panel) in the same slice as in Figure \ref{fig:slice-reionization}.%
}
\end{center}
\end{figure*}

\subsection{Preparing numerical simulation}
\label{subsec:preparing_numerical_simulation}

For the following analysis we use the down-sampled snapshots of $40 \hMpc$ to $256^3$ uniform grid (the effective resolution is $156 \hkpc$). Next, for each pixel we calculate the moment of its ionization. We adopt two thresholds 10\% and 90\% weighted by mass. The vast majority of the cells cross these thresholds only once, therefore we can assign those values to the grid. In Figure \ref{fig:slice-reionization} the slices of the redshift of ionization are plotted. In Figure \ref{fig:slice-delta} the same slice for the density initial conditions (IC) and the evolved baryonic density field at $z=6$ are plotted.

Immediately we see that the ionization field is much smoother for 10\% threshold rather than for 90\% threshold. It motivates us to consider the time delay between the moments when a cell reaches 10\% and 90\% ionization thresholds as a separate field.

\begin{figure}
\begin{center}
\includegraphics[width=0.95\columnwidth]{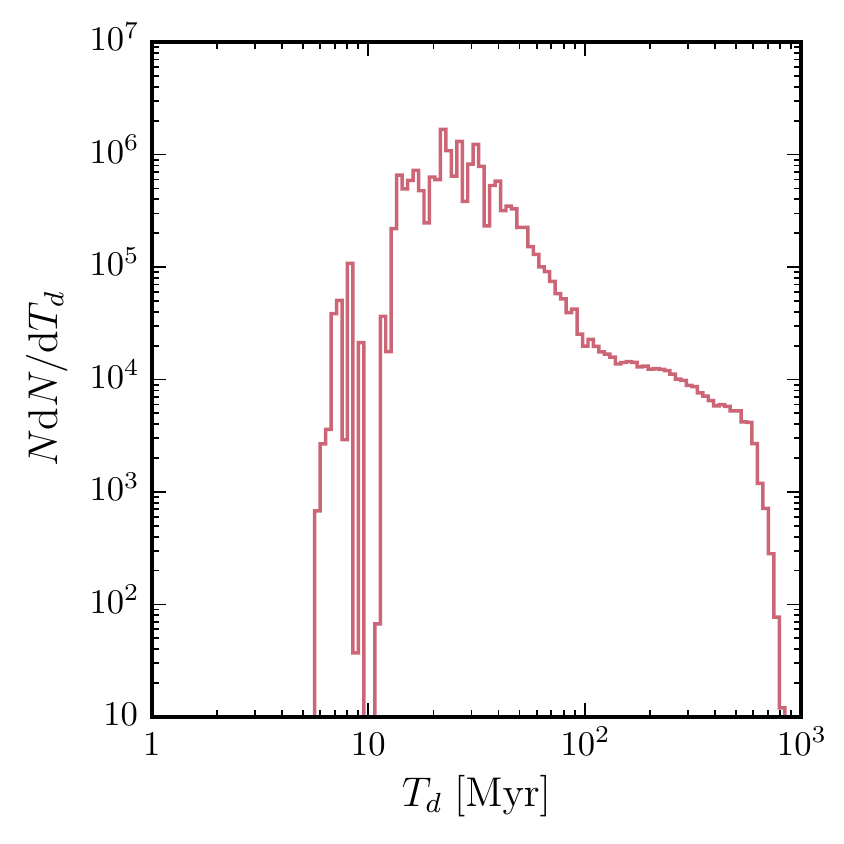}
\caption{\label{fig:time_delay_hist} The distribution of time delays between the moments when a cell reaches 10\% and 90\% ionization thresholds. The spikes at $T_d \sim 8$ and the discontinuity at $T_d \sim 10$ are artificial, and are caused by the uneven time intervals between the snapshots.%
}
\end{center}
\end{figure}

In Figure \ref{fig:time_delay_hist} we plot the distribution of those delays. For the vast majority of the cells this delay is small compared to the cosmological timescales; however, a fraction of the cells show big delays. The peak-like structure of the histogram is an artifact due to the characteristic time between the simulation snapshots.

\begin{figure}
\begin{center}
\includegraphics[width=0.95\columnwidth]{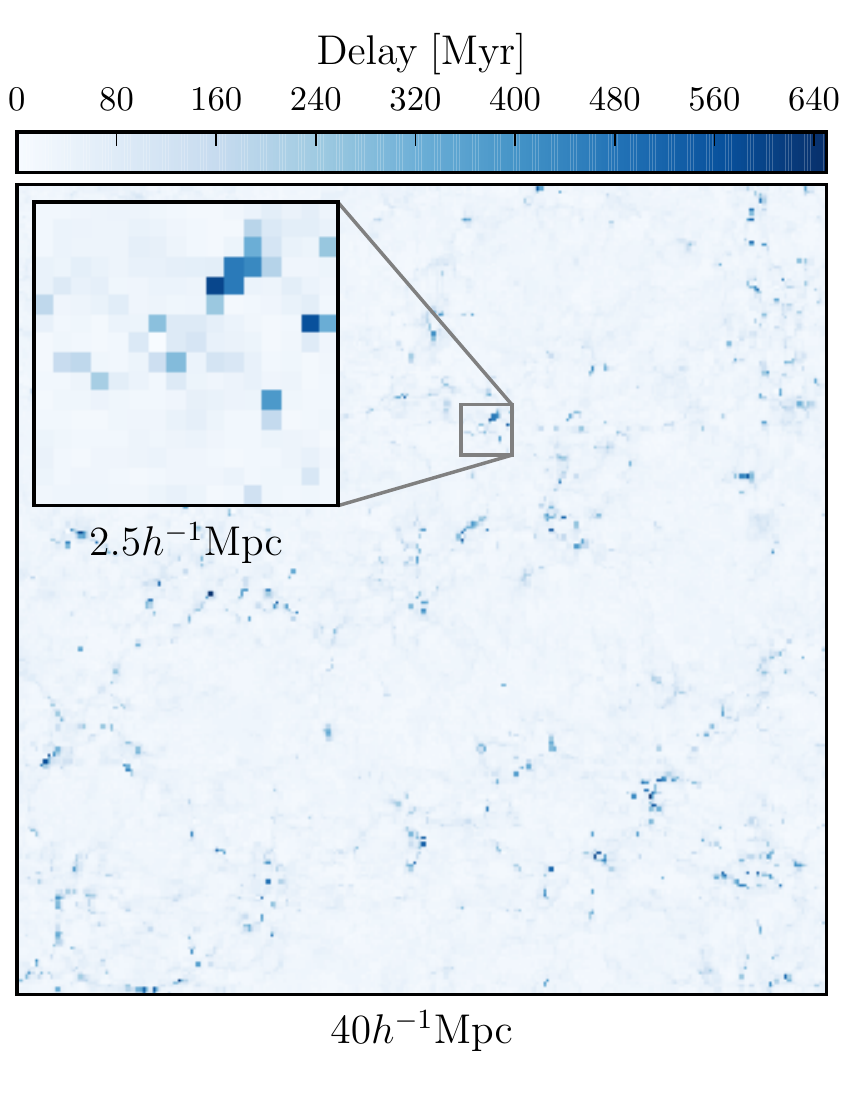}
\caption{\label{fig:time_delay_slice} The time delay between the moments when cell reaches 10\% and 90\% ionization thresholds in the same slice as in Figures \ref{fig:slice-reionization} and \ref{fig:slice-delta}. %
}
\end{center}
\end{figure}

In Figure \ref{fig:time_delay_slice} we plot the delay for the same slice. The cells with long delays are correlated with the overdense cells in the density field. This correlation is caused by the filaments located in denser cells, which stay relatively neutral even when the surroundings are already highly ionized. Therefore, this effect fits into the MHR00 paradigm of the gradual ionization of the overdense regions.

\subsection{Random walk in numerical simulation}

The trajectory, or random walk, which we described in \S\ref{subsec:FZH04}, can be also defined in a numerical approach, i.e. in the simulation box. The smoothing of the scalar field can be performed in the same manner through Fourier space, applying top-hat filter. The main difference is that the trajectory is not longer a continuous function, and therefore we have to discretize it. We choose a number of the physical scales $r_0, r_1, ... r_N$ in comoving units. 

In this paper we stick with $r$ notation since we work with different scalar fields, and values of $m$ and $\sigma^2$ are not constant across them. Also not all scalar fields we use are Gaussian, and those are not described exclusively by the first moment statistics, power spectrum. Therefore, $\sigma^2$ is less motivated quantity.

We call the \textit{barrier} a two dimensional function, $B(r_i,z_j)$, which is defined on uniform grid of scales, $r_i$, and redshifts, $z_j$. The \textit{trajectory} is one dimensional function associated with each cell $x$, $T_x(r_i)$, defined at all scales, $r_i$. It corresponds to the density of scalar field in the cell after applying a smoothing filter of scale $r_i$.     

% A large variety of filters exist. The most physically motivated (based on the analytical model of reionization) is a spherical filter. Other filters, like the Gaussian filter, do not correspond to a meaningful shape in the real space, while the spherical filter is a good approximation of ionization front from a single source.       

The analytic theories based on \cite{Furlanetto_2004} use the first time crossing as an indication of ionization. The condition of the first time ionization of a cell $\vec{x}$ can formulated as follows:   
\begin{equation}   
z_{ion} = \max \{ z_j \; | \;\exists i.\{T_x(r_i) > B(r_i,z_j)\}.
\end{equation}
It summarizes the barrier approach.

In order to translate a numerical simulation to the language of the analytical model we calculate a random walk for every cell in our simulation box. In order to do this we use 8 log-spaced discrete smoothing scales -- from a single cell to the half-box size. This number is found to be sufficient for our purposes, and increasing it does not affect the results. We generate a smoothed density field with every chosen scale. Then, we normalize the resulting scalar fields to the Normal distribution with zero mean and unity standard deviation. This operation is performed only to avoid usage of physical units (which differ from field to field) and better visualization of trajectories in figures. Skipping this step does not affect the final results. Finally, we assign to each cell 8 numbers that correspond to the normalized overdensities for 8 smoothing scales.

The method described above can work with any scalar field. In this paper we consider only the IC density field (linear) and evolved matter density field (non-linear). We leave the star density and halo density fields beyond the scope of this paper; however, both of these fields potentially can significantly improve the results.

\begin{figure}[h!]
\begin{center}
\includegraphics[width=0.95\columnwidth]{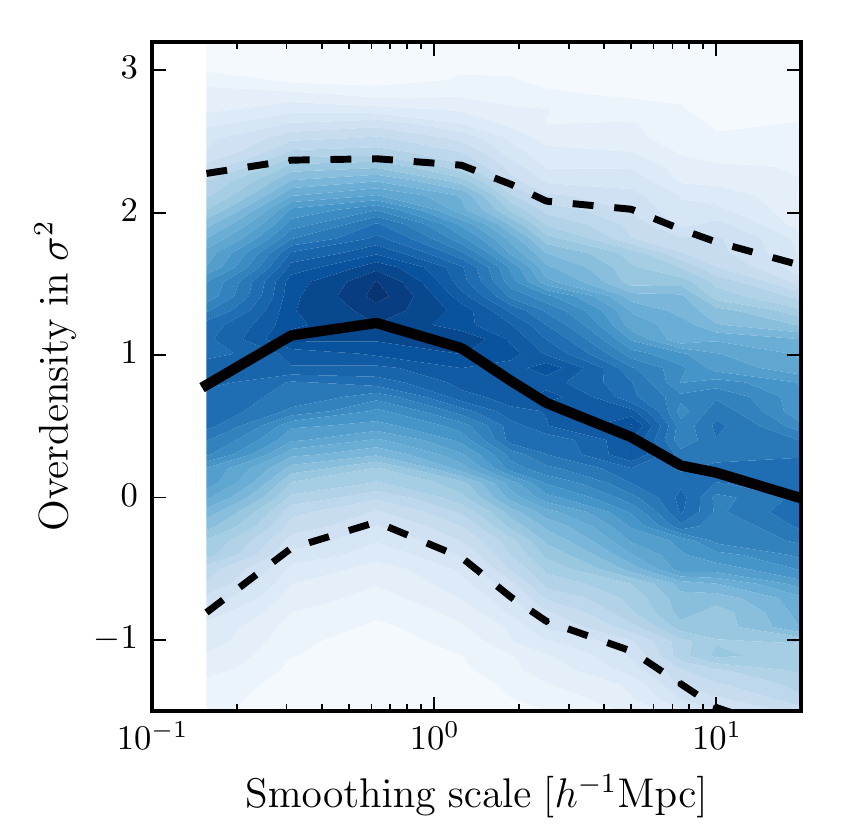}
\caption{\label{fig:random_walk_dens} The density of the trajectories distribution. The solid and dashed lines are the median trajectory and $1\sigma$ scatter.
}
\end{center}
\end{figure}

For now, let us consider the IC, which is a Gaussian field, and the redshift of 10\% ionization. We want to determine whether there is a correlation between the trajectories and the redshifts of ionization. We consider the cells that were ionized within some redshift interval and plot the density distribution of their trajectories along with the median in Figure \ref{fig:random_walk_dens}. The deviation from zero tells us that, indeed, there is a dependence.

Next, we repeat this operation for all redshift intervals. We consider two density fields (IC and evolved DM density) and two ionization thresholds (10\% and 90\%). Thus, we have 4 possible combinations. The median trajectories for all of them are plotted in Figure \ref{fig:median_traj}. Also, for completeness, we study delays with the same approach and evolved density field (see Figure \ref{fig:random_walk_delay}).

\subsection{Discussion}
\label{subsec:discussion}

\begin{figure*}
\begin{center}
\includegraphics[width=0.9\columnwidth]{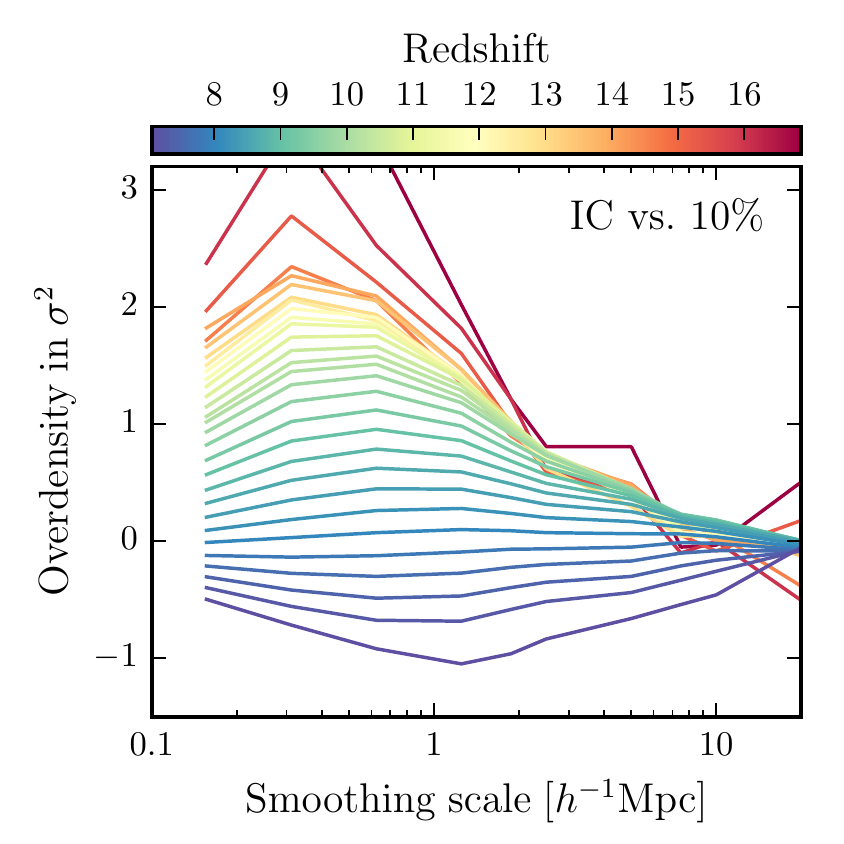}$\;\;\;\;\;\;\;$
\includegraphics[width=0.9\columnwidth]{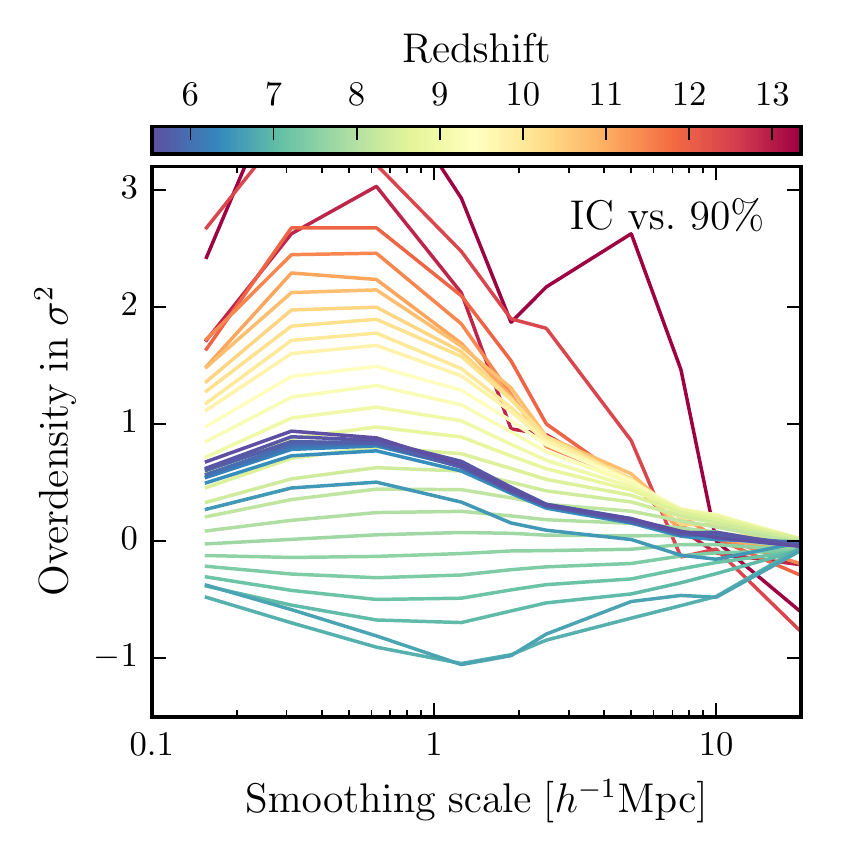}
\linebreak

\includegraphics[width=0.9\columnwidth]{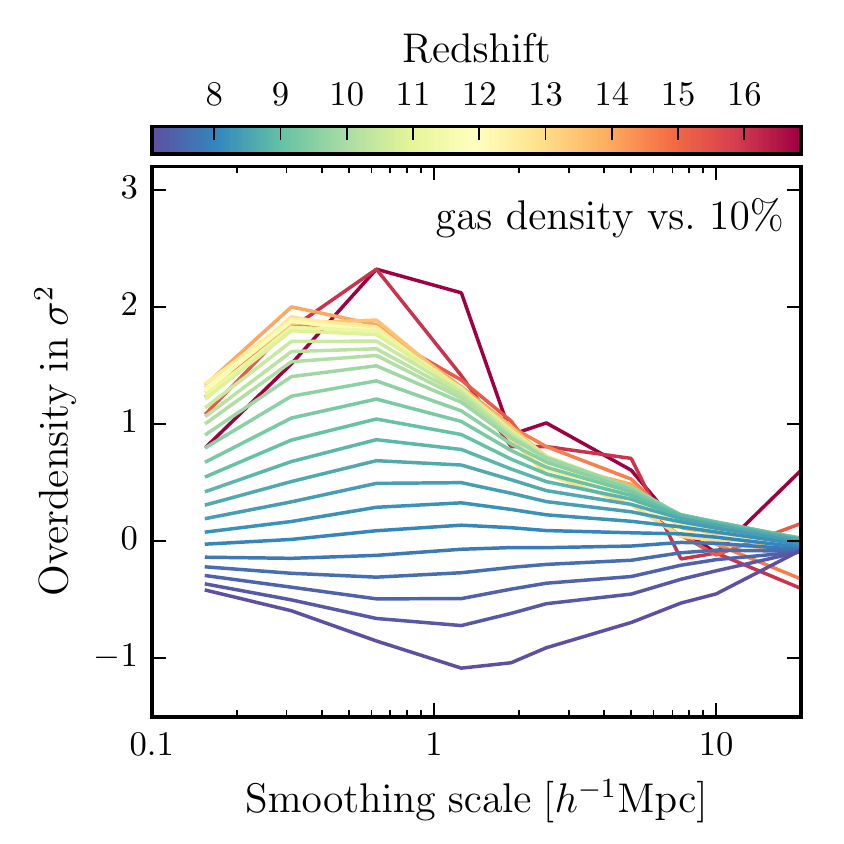}$\;\;\;\;\;\;\;$
\includegraphics[width=0.9\columnwidth]{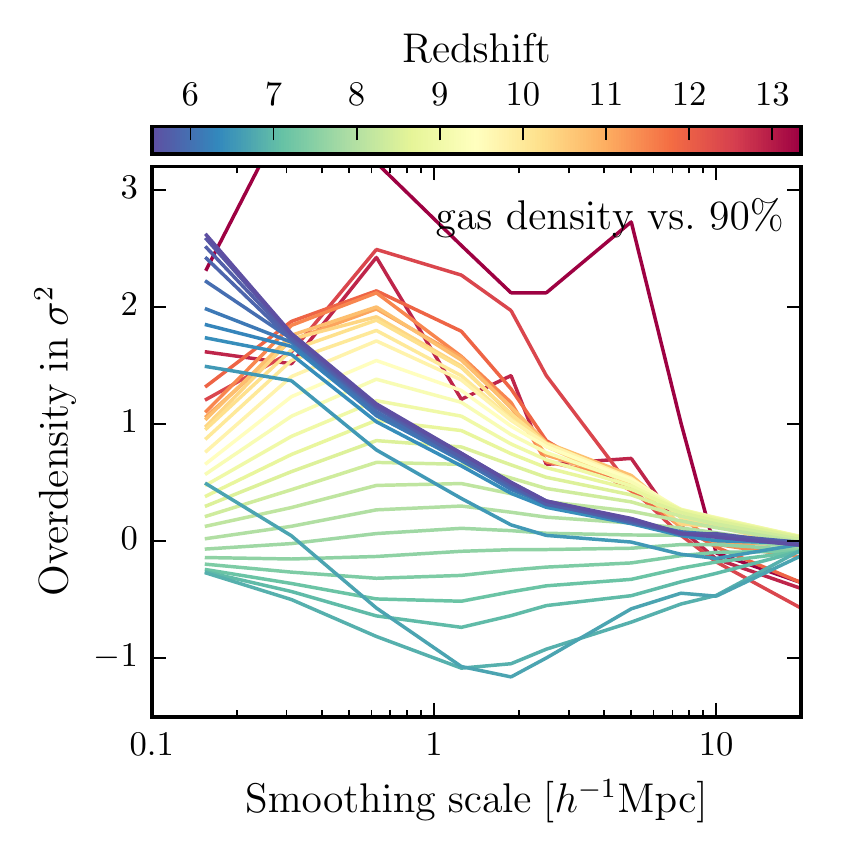}
\caption{\label{fig:median_traj}The phase space of median trajectories colorcoded with redshift and defined for IC and 10\% ionized threshold ionization field (top-left panel); IC and 90\% (top-right panel); evolved density field and 10\% (bottom left); evolved density and 90\% (bottom right).%
}
\end{center}
\end{figure*}

\begin{figure}
\begin{center}
\includegraphics[width=0.95\columnwidth]{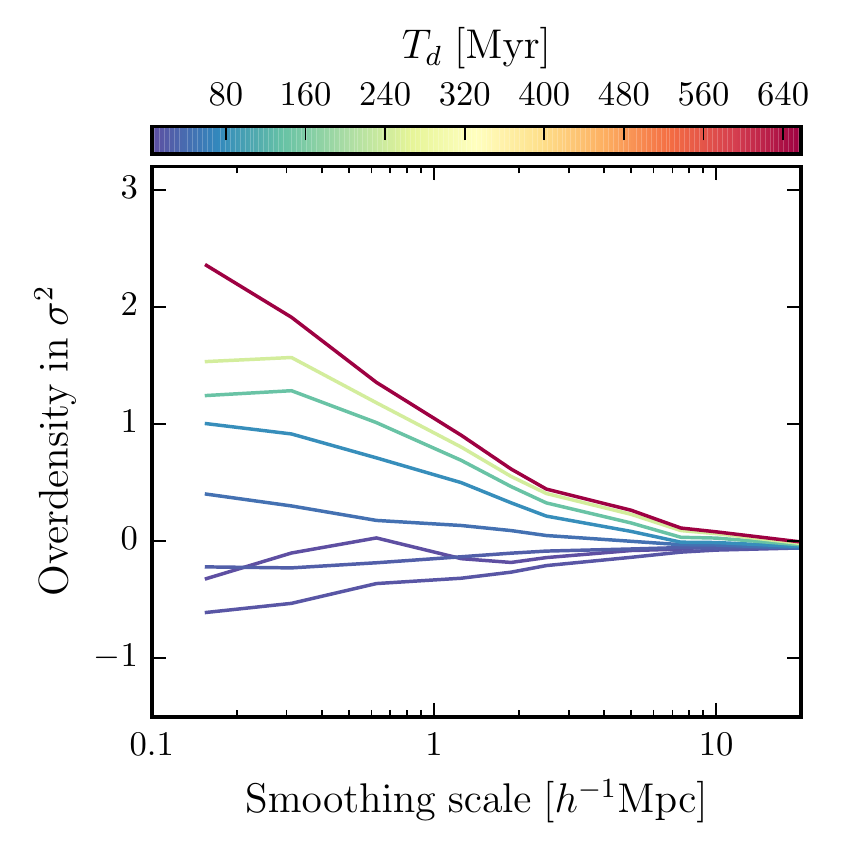}
\caption{\label{fig:random_walk_delay}Median trajectories corresponding to the delay between crossing 10\% and 90\% ionization thresholds.%
}
\end{center}
\end{figure}

The median trajectories in Figures \ref{fig:median_traj} and \ref{fig:random_walk_delay} clearly show redshift dependence. In this section we discuss the observed patterns and the physical reasoning behind them.

First, we have two figures which relate to the 10\% ionization threshold with the IC and the evolved density fields. Both of them display a gradual decrease of a barrier from overdense to underdense, which confirms the inside-out scenario. The physical explanation for this observation is simple: the dense regions host galaxies which produce the ionizing radiation, and therefore reionization starts from the overdense regions and ends inside the underdense -- voids.

All the trajectories reach zero at the maximum smoothing scale which is expected box effect; however, the same behavior is expected in the real world, since cosmic variance is supposed to reach zero at some scale. Since the largest box we have access to is $40\hMpc$, we cannot distinguish these two effects. %We will be able to draw a more specific conclusion when 80$h^{-1}$Mpc will be done.

On the other end, at the smallest scales, there is also a reduced correlation between density and ionization field. For the IC density it is caused by the fact that the position of galaxies is not identically correlated with density peaks in the IC. For the evolved density field the reasoning is different. The evolved density field correlates with galaxy positions much better; however, it also correlates with dense filaments which require more intense flux of ionizing photons to ionize.

The phase plots show that the highest correlation (largest amplitudes of the trajectories) is achieved at scales $0.5-2\hMpc$. This might be interpreted as the range of scales where the majority of the information about the ionization field is contained. As mentioned above, due to the lack of larger boxes we can only speculate regarding the upper bound of such a range. However, we can be more confident in the lower bound because the resolution of the numerical simulation ($0.6\hkpc$) is much higher than the down-sampled grids ($156\hkpc$) that we used for this analysis.

% In the performed analysis we only used one resolution, $156\hkpc$, which, in our experience, is the most optimal for highlighting all of the main features. Refining the resolution would allow one to extend the phase plots to smaller scales. These scales would allow to resolve individual filaments and therefore more accurately predict their neutral fraction. However, resolving galactic scales and ISM is not useful. For the large scale morphology of ionization fronts, the scales smaller than $100-200\hkpc$ do not matter. 

Another feature that can be observed is the scale of the largest correlation amplitude. It increases from just below $1\hMpc$ at redshift ~15 to a few $\hMpc$ scale at redshift ~6. It can be associated with the growth of the characteristic bubbles size.

The most important observation from these phase space plots is that the median trajectories are distinguishable and therefore they contain information about the reonization history. It confirms the applicability of the excursion set framework for describing the reionization morphology.

Now we consider the figures related to 90\% threshold. The main difference is located at lower redshifts, where the median trajectories are going up again. Those correspond to the regions correlated with the local density which we mentioned in \S\ref{subsec:preparing_numerical_simulation}. The IC density phase plot shows weaker upturn than evolved density because, again, the density peaks in the IC are less correlated with filaments than in the evolved field.

This upturn reveals the transition between two regimes FZH04 and MHR00. By design the MHR00 model describes the late stages of reionization. It shows that denser regions will be ionized at later times. It is what we can see as growing trajectories in the phase plot. The scale of the largest correlation for this regime is the smallest in our analysis.

The question is if the trajectories can be associated with the redshift of ionization one to one. We see in Figure \ref{fig:median_traj} that for 90\% ionization threshold some trajectories are indistinguishable from one another. However, the phase space of delayed field in Figure \ref{fig:random_walk_delay} does not have intersecting trajectories. It means that by splitting 90\% ionization threshold field into 10\% ionization field and delays we can avoid degeneracy in the phase space.

Another way of thinking about it is that MHR00 model is valid not only at late stages of reionization, but at any redshifts inside the ionized bubbles. While the FZH04 model only describes the ionization fronts at low ionization thresholds (10\% in our case).

\begin{figure}
\begin{center}
\includegraphics[width=0.95\columnwidth]{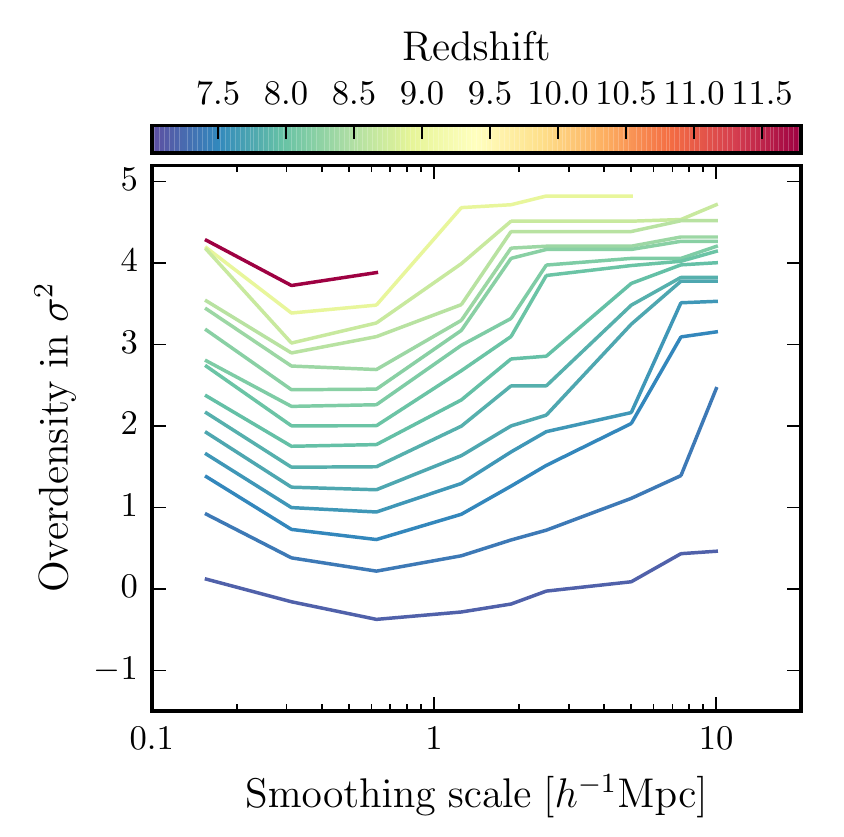}
\caption{\label{fig:estimated_barriers}The barriers that were estimated from the phase space, shown in Figure \ref{fig:random_walk_dens}.%
}
\end{center}
\end{figure}

%----------------------
\section{Training the analytical model on a numerical simulation}
\label{sec:tuning}
%----------------------

In \S\ref{sec:analitic_for_numerical} we applied the framework of the analytical model (the excursion set formalism) to the numerical result. The measured trajectories behave as expected, which makes us more confident about the analytical approach. Hence, in this section we study how well the analytical model can match a simulation.

% \cite{Majumdar2014}

One could try to construct an analytical model which would incorporate the physics identical to the numerical simulation, and then compare both results. Instead, we consider a different approach. We question how well the analytical model can describe a given ionization field without relying on the underlying physics. We perform a sort of reverse engineering of numerical result, trying to find the barrier in analytical model, which can produce this ionization field.

Then in \S\ref{subsec:combining_models} we describe our method of training analytical model and 
in \S\ref{subsec:barrier_estimation} we describe the method of extracting a barrier from phase space.

The question which immediately emerges when we talk about training or fitting is what is the criteria of goodness. We overview them in \S\ref{subsec:tuning_criteria}.

%----------------------
\subsection{Building a model}
\label{subsec:combining_models}
%----------------------

As it was shown in \S\ref{sec:analitic_for_numerical} there is a non-negligible delay between a pixel approaching 10\% and 90\% ionization levels. While the trajectories, which correspond to 10\% ionization level, are distinct, those for 90\% level intersect with each other. It means that the random walk by itself is not informative enough to predict 90\% ionization in some regions. However, it was also shown in \S\ref{sec:analitic_for_numerical} that the median trajectories which correspond to the delay are distinguishable, and therefore can be described by a trajectory.

Thus, we propose to consider the ionization field as consisting of two components: the redshift of crossing 10\% ionization threshold and the delay to the 90\% threshold crossing. Both of these fields are well distinguishable in the trajectory phase space. 

These two components can be considered as a combination of the FZH04 and the MHR00 models. The FZH04 model describes morphology of ionized patches at large scales while the MHR00 is responsible for neutral patches inside ionized volume which take longer time to ionize.

%----------------------
\subsection{Barrier estimation}
\label{subsec:barrier_estimation}
%----------------------

In \S\ref{sec:analitic_for_numerical} we operate with median trajectories; however, the excursion set formalism model, the FZH04 model, relies on barriers. Nevertheless, the median trajectories are useful because they are easy to define and give a good sense of the overall behavior of the trajectories. 

Now if we imagine a simulation generated using excursion set formalism, then it would be possible to derive the barrier by simply calculating a $\sim$99.9\% percentile line, instead of the median in Figure \ref{fig:random_walk_dens}. This will work for analytical model, because all of the trajectories that were ionized in the redshift interval [$z_i$, $z_{i+1}$] cross the barrier $B(z_{i+1}, r)$ and do not cross $B(z_i, r)$. Since it is not the case for a real numerical simulation with radiative transfer, because we have at least some noise in the phase plot, we need to come up with a more robust algorithm.

In Figure \ref{fig:random_walk_dens} the two-dimensional density of trajectories is plotted for a given redshift. In fact, the phase plot is three-dimensional, where the third dimension is the redshift. In this space the series of barriers at different redshifts is represented by a surface. 

The simplest algorithm would be the brute-force, i.e. to probe all possible barriers and find the one that allows to achieve the best concordance with the reference simulation. This would be the most robust, but not effective algorithm. We can optimize it by making an initial guess of a good barrier, and then only brute-force all barriers close to it. %The algorithm that helps us to make an initial guess is described in Appendix \ref{app:barrierest}, and based on the  phase space shown in Figure \ref{fig:median_traj}. 

In order to make a good initial guess for the barrier we consider the gradients along redshift axis. The collection of points with highest gradients defines the surface, which is our best guess. Notice, that for the scale equal to our box size all gradients along the redshift axis will be zero, because the trajectories are not distinguishable on this scale. Consequently, the maximum gradient cannot be found and the barrier is not defined at this scale (see Figure \ref{fig:estimated_barriers}).

No matter what is the training algorithm, it should rely on some sort of \textit{goodness} criteria, which we discuss in \S\ref{subsec:tuning_criteria}. Here we choose to optimize the cross correlation coefficient (see Equation \ref{eq:rxd}) on all scales. The resulting barriers for the given simulation and 10\% ionization threshold are presented in Figure \ref{fig:estimated_barriers}. Depending on the application, one can choose a different criteria, for instance limit the scales of interest, and have a slightly different best fit.

The shape of the barriers can be qualitatively explained as follows. At the largest scales all median trajectories are indistinguishable, therefore there is not much information at those scales in the phase space and the barrier approach is not effective. Thus, the barrier should be high at those scales in order to have negligible effect. Same logic can be applied for the smallest scales where median trajectories for 10\% ionization threshold also approach zero. The majority of information is contained at intermediate scales, therefore in order to capture it the barrier has to be lower in the middle.

\begin{figure*}
\begin{center}
\includegraphics[width=.66\columnwidth]{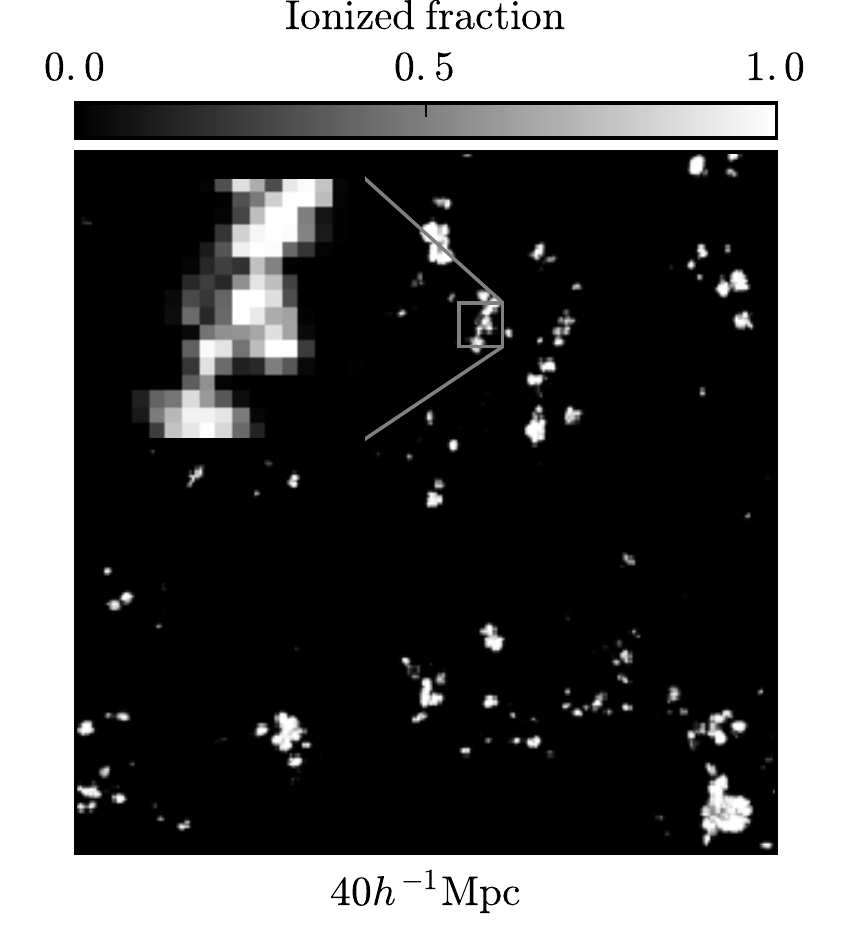}
\includegraphics[width=.66\columnwidth]{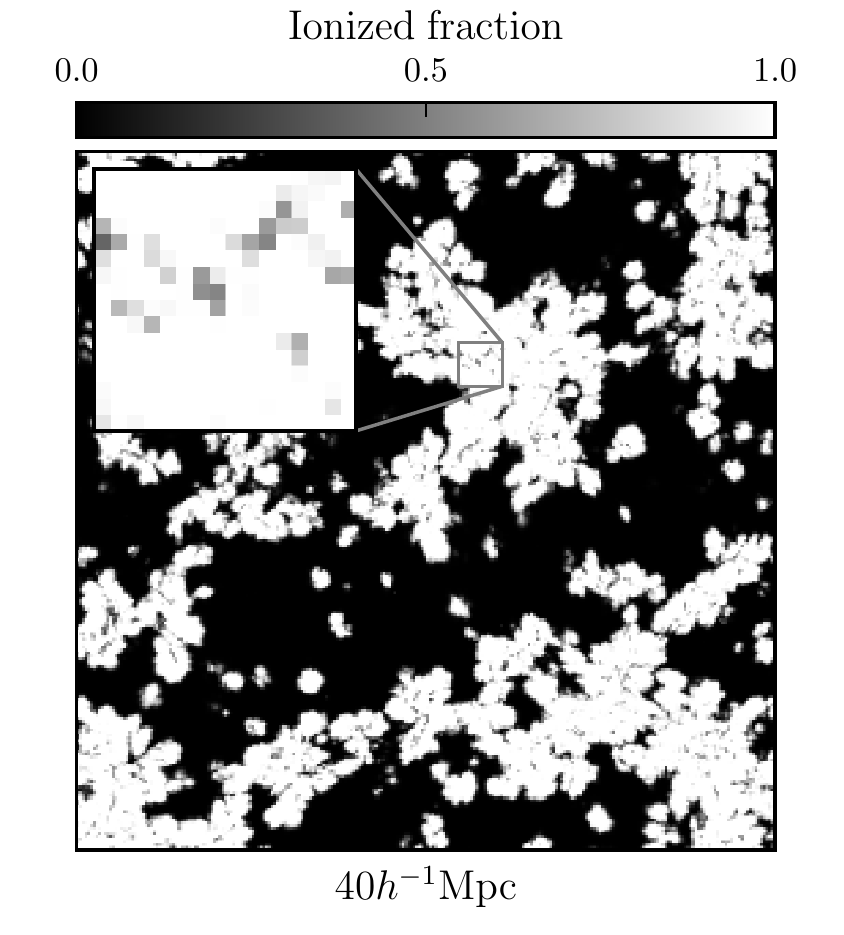}
\includegraphics[width=.66\columnwidth]{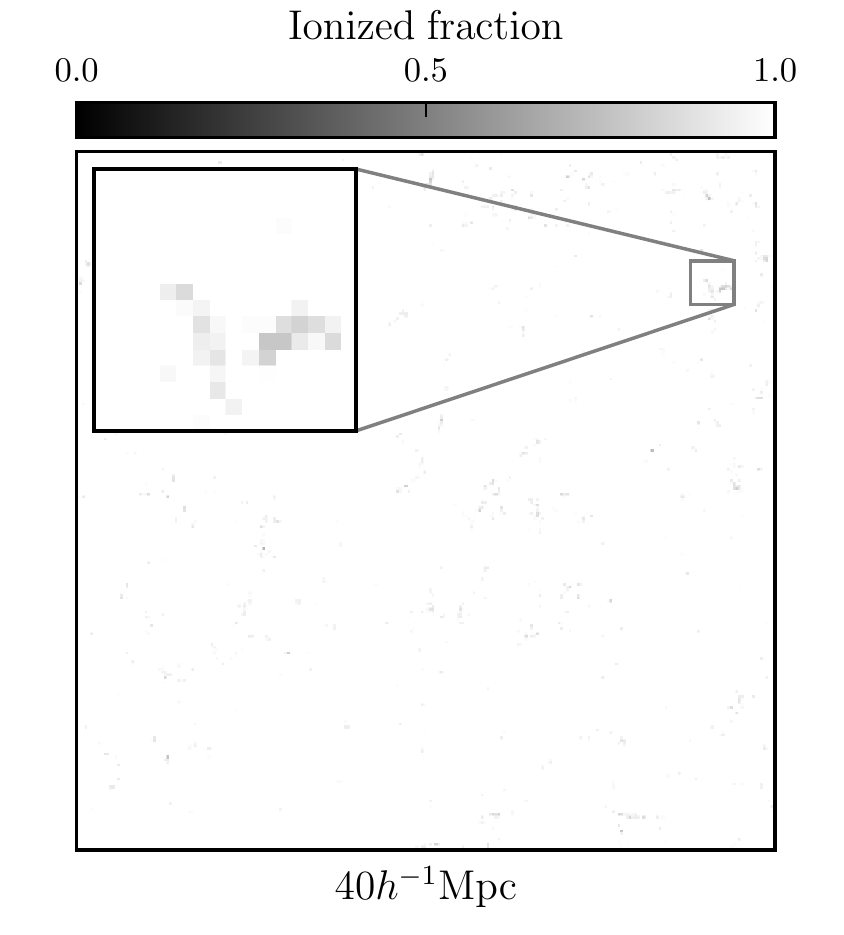}
\includegraphics[width=.66\columnwidth]{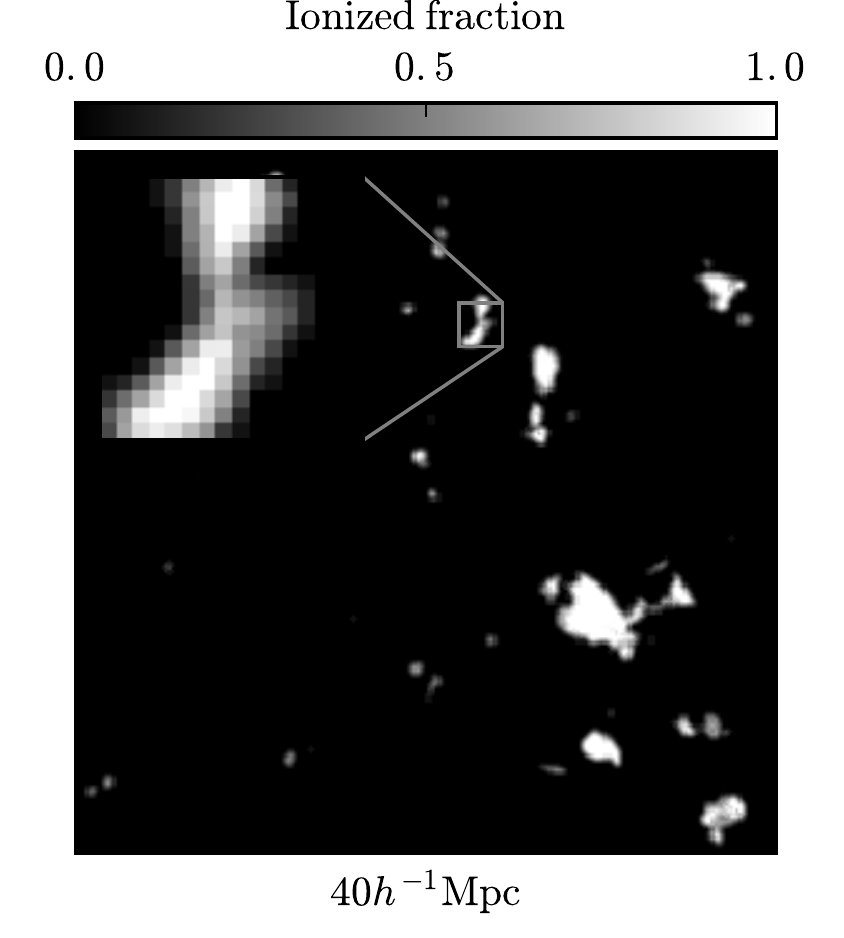}
\includegraphics[width=.66\columnwidth]{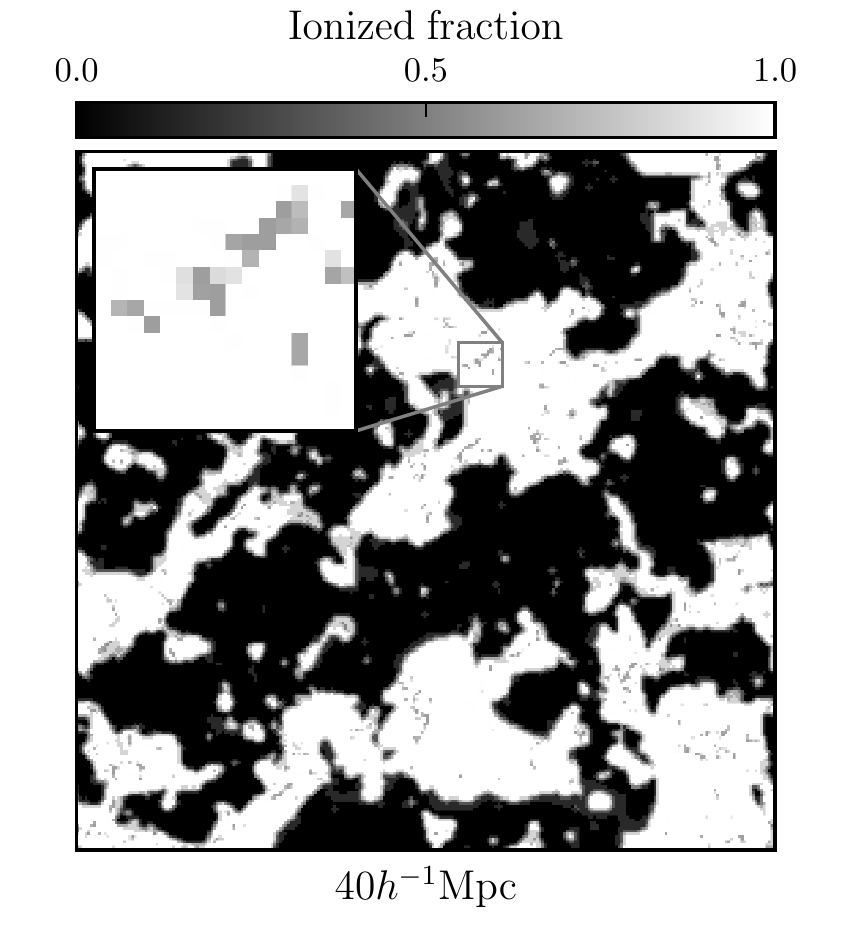}
\includegraphics[width=.66\columnwidth]{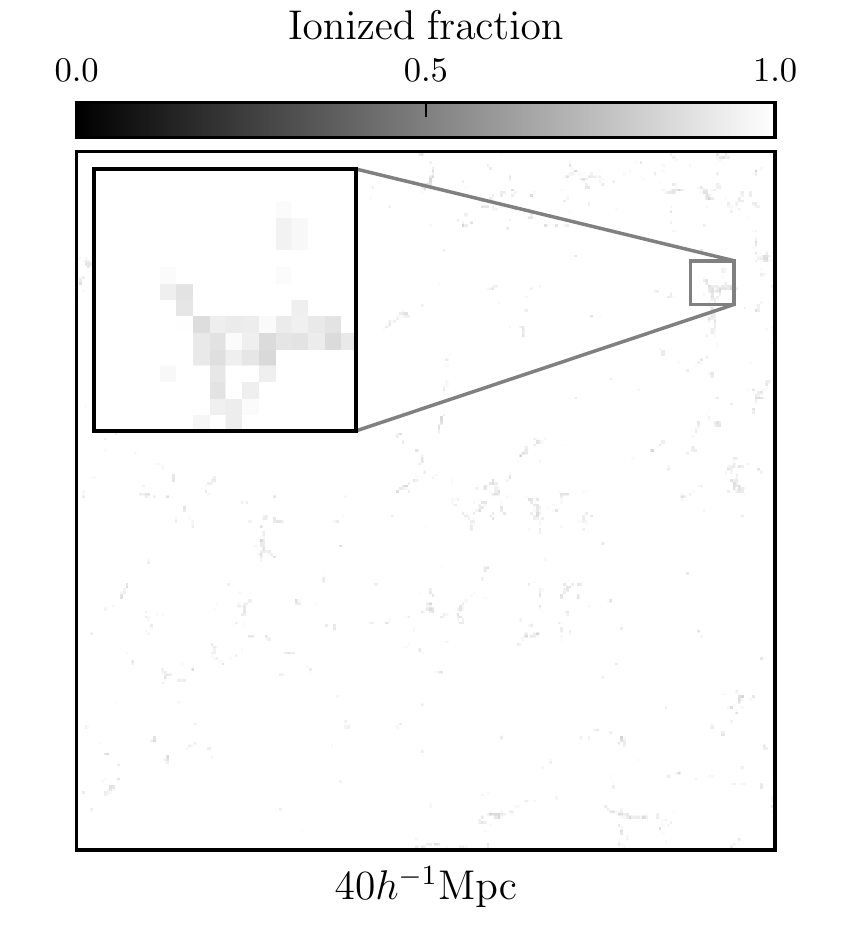}
\includegraphics[width=0.66\columnwidth]{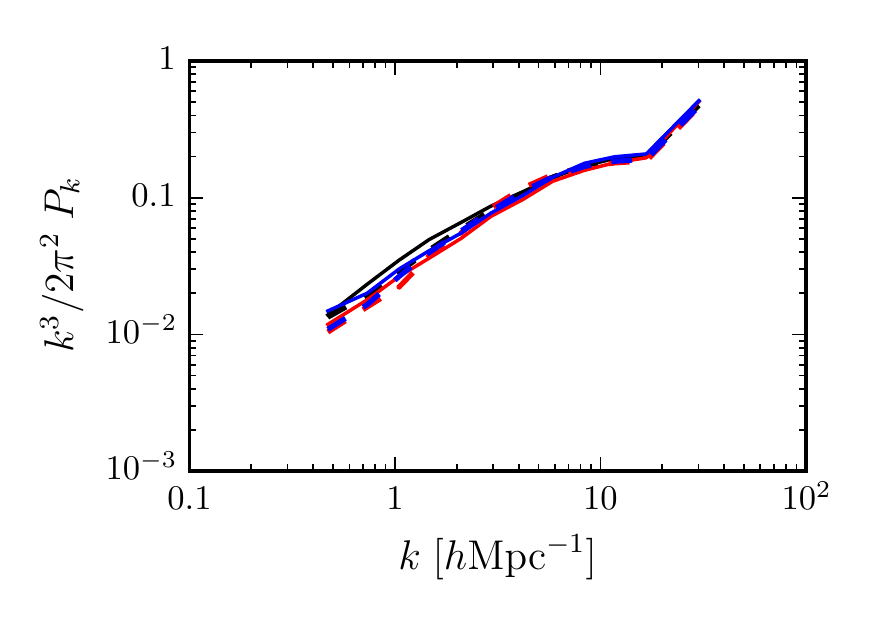}
\includegraphics[width=0.66\columnwidth]{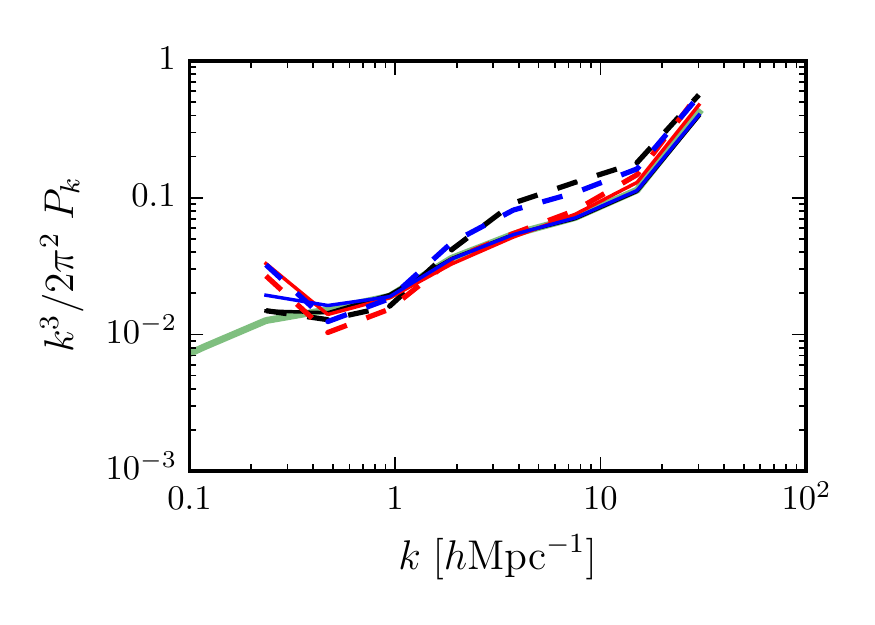}
\includegraphics[width=0.66\columnwidth]{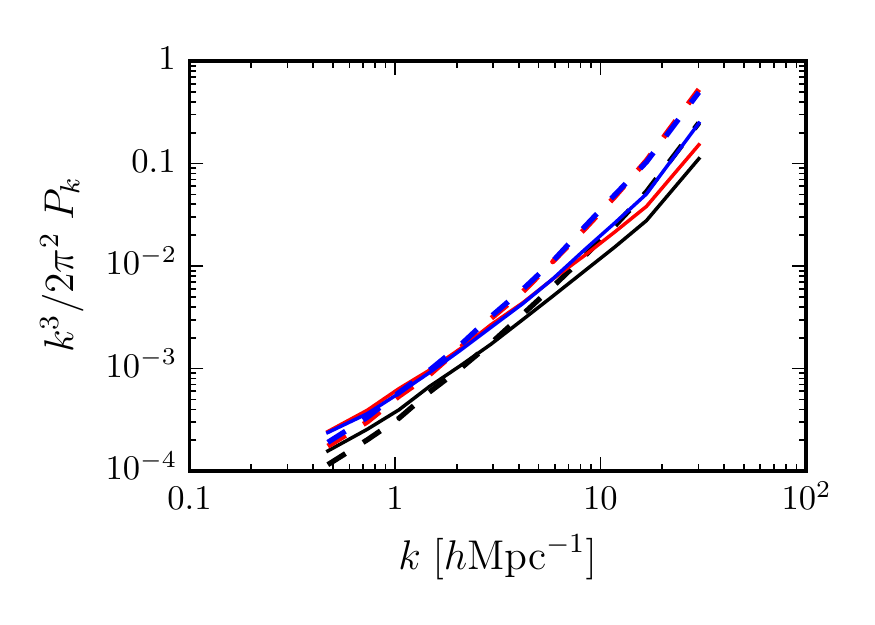}
\includegraphics[width=0.66\columnwidth]{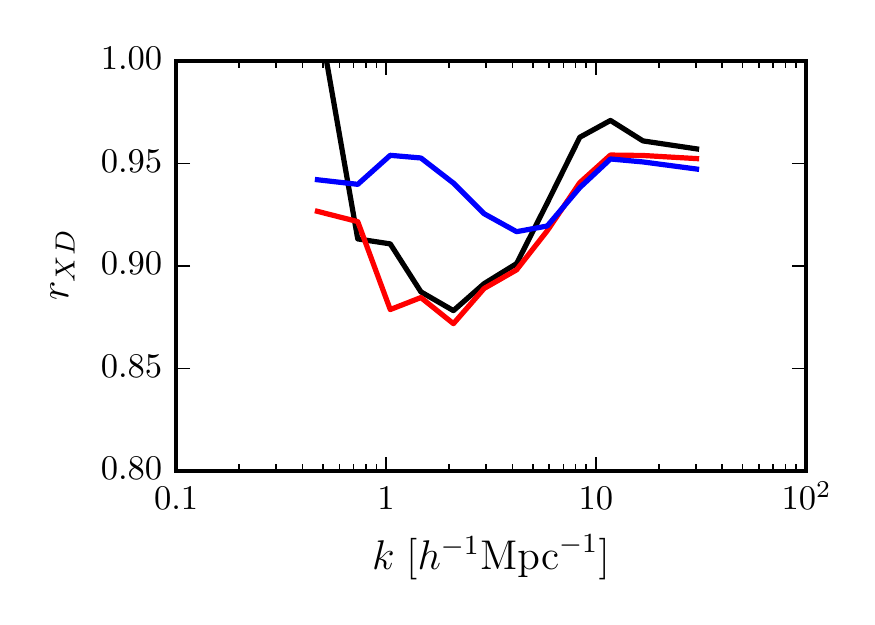}
\includegraphics[width=0.66\columnwidth]{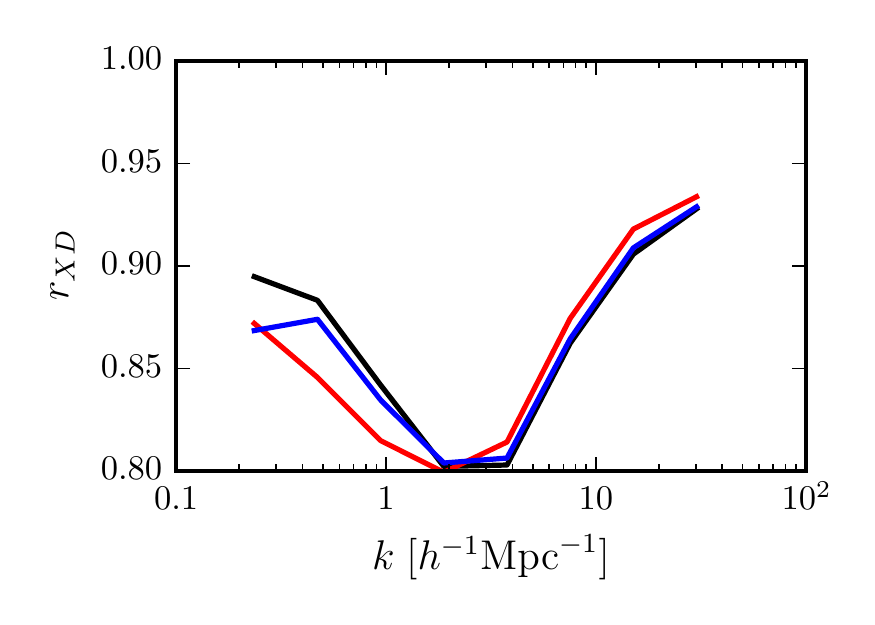}
\includegraphics[width=0.66\columnwidth]{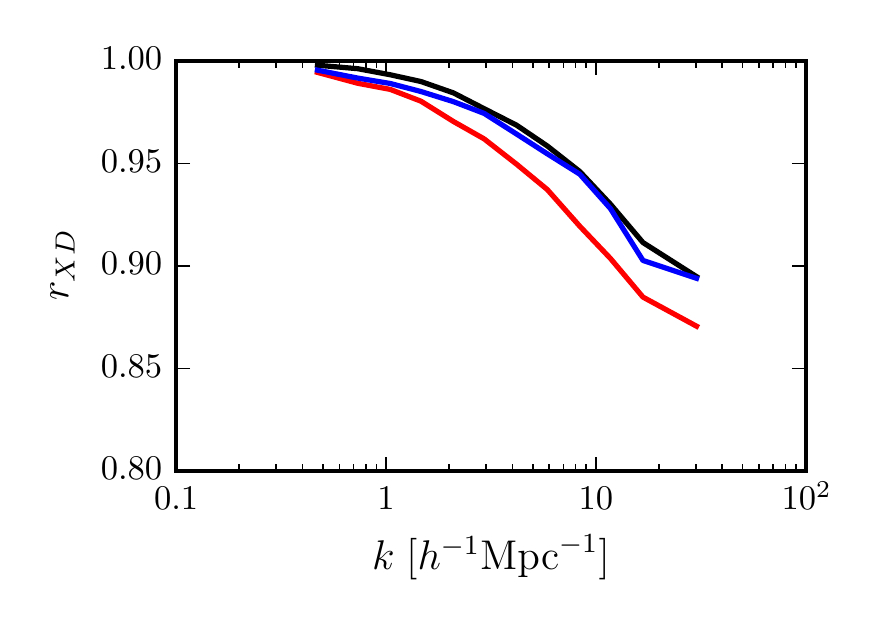}
\caption{\label{fig:simmodel}
The ionization fraction in the numerical simulation (the first row of panels) and trained analytical model (the second row) in the same slice as in Figures \ref{fig:slice-reionization}, \ref{fig:slice-delta}, and \ref{fig:time_delay_slice}. The power spectrum of the neutral hydrogen in the numerical simulations (dashed lines) and our model (solid) is shown in the third row of panels. The correlation coefficient (see Equation \ref{eq:rxd}) is shown in the forth row.  Black is the realization A, blue and red are realization B and C. Solid lines correspond to the same analytical model, which is fitted only to the realization A. Green solid line corresponds to the power spectrum calculated with the same analytical model in a 160$\hMpc$ box.
The columns correspond to three different moments in time: early stage of reionization ~5\% ionized fraction; intermediate stage with ~50\% ionized fraction and interloping bubbles; and late stage, when only filaments remain partially neutral. %
}
\end{center}
\end{figure*}

The slices of the numerical simulation and the result of our trained analytical model at redshift 8 is shown in Figure \ref{fig:simmodel}. The model is trained to match the power spectrum of neutral hydrogen;  however, even visually we can see that analytical model mimics the large scale structure of ionization fronts. 
 
%\begin{figure*}
%\begin{center}
%\includegraphics[width=1.\columnwidth]{figures/real}
%\includegraphics[width=1.\columnwidth]{figures/model}
%\caption{\label{fig:simmodel}
%The ionization fraction in the numerical simulation (left panel) and trained analytical model (right panel) in the same slice as in Figures \ref{fig:slice-reionization}, \ref{fig:slice-delta}, and \ref{fig:time_delay_slice}. The total ionization fraction ~50\%. Zoom-in region (2.5$\hMpc$) shows that the analytical model reproduces semi-neutral substructure on small scales.%
%}
%\end{center}
%\end{figure*}
%
%\begin{figure}
%\begin{center}
%\includegraphics[width=0.95\columnwidth]{figures/Pk_updated}
%\caption{\label{fig:pkfits}
%Dashed lines are the actual power spectrum of neutral hydrogen at redshift 8 (~50\% of global ionization fraction) measured in the simulation boxes. Black is the realization A, blue and red are realization B and C. Solid lines correspond to the same analytical model, which is fitted only to the realization A. Green solid line corresponds to the power spectrum calculated with the same analytical model in a 160$\hMpc$ box.%
%}
%\end{center}
%\end{figure}
%
%\begin{figure}
%\begin{center}
%\includegraphics[width=0.95\columnwidth]{figures/rxx_updated}
%\caption{\label{fig:rxx}
%Cross correlation coefficients between the neutral hydrogen in the numerical simulation and the analytical model at the same moment as shown in the Figure \ref{fig:pkfits}. Black is the realization A, blue and red are realization B and C.%
%}
%\end{center}
%\end{figure}

%----------------------
\subsection{Goodness criteria}
\label{subsec:tuning_criteria}
%----------------------

The comparison between the numerical and analytical models had been carried out before in \citet{Zahn2011} and \citet{Battaglia2013}. Here we outline possible criteria and discuss why some of them are not applicable in our case.

\textit{Global ionization fraction}, or optical depth averaged over the whole sky, does not trace any information about morphology and inhomogeneity of reionization. Therefore, it can be used only for estimating total number of emitting photons, neglecting recombination (which in a general case can significantly depend on the morphology).

\textit{Bubble distribution} has a few definitions. The first one is robust and based on smoothed scalar fields. It is equivalent to the power spectrum. On the other hand, if a bubble is defined as an ionized volume, which is not connected with other bubbles through ionized patches, than it becomes much harder to define it algorithmically. If the bubbles are defined as the largest spherical volume with some given ionized fraction \citep{Zahn2007}, then in our case it is not clear how to define bubbles for the dense semi-neutral filaments and for the cells right near them.

\textit{Pixel-by-pixel} comparison, which was made in \citet{Battaglia2013}, is the most powerful in case if we expect perfect match. However, if we expect some deviations we will face a few uncertainties. First, as we mentioned before the redshift of ionization of a cell is not a well defined quantity. It depends on the threshold, and also the process of ionization cannot be instantaneous. Second, one should ask which areas are more important to match. For instance, at redshifts below 6 most of the observed signal comes from neutral patches which occupy small volume; therefore, pixel-by-pixel comparison has to take it into account weighting pixels by their contribution to the observed quantities.

\textit{Topological quantities}, such as Minkowski functional are used for statistical analysis of reionization and distribution of cosmological objects \citep{Friedrich2011, McDonough2013}. However, these quantities are not directly observed.

\textit{Power spectrum} is the most common statistics for describing fields on large scales. Also it is a direct observable which will be measured with 21cm experiments. In a vanilla FZH04 model, where only excursion set formalism and the Gaussian IC are used the barrier and the power spectrum (as well as bubble distribution) are directly interconnected and one can be derived analytically from another. However, when we modify the model by using non-linear density field and adding the neutral patches inside the ionized regions\footnote{The input of smallest structures to the power spectrum has been studied in \citet{Watkinson2015}. Also, in \citet{2015arXiv151008767K} the effect of the substructure on the 21 cm power spectrum is studied.} this convenience vanishes and we have to perform all calculations numerically.

\textit{Cross correlation} coefficient between the neutral hydrogen in the numerical simulation, $D$, and analytical model, $X$, defined as \citep{Zahn2011}: 
\begin{equation}
\label{eq:rxd}
r_{XD}=P_{XD}/\sqrt{P_{XX}P_{DD}}.
\end{equation}

In this study we consider only last two statistics because those are directly related to the observable quantities and do not have uncertainties in the definition. The result of training is shown in the Figure \ref{fig:simmodel} with black lines. The analytical model matches the numerical simulation within 10s of percent level, which is consistent with the similar study in \citet{Zahn2011}. In contrast to that study, we consider smaller scales and semi-neutral filaments, and in result we have a different shape of $r_{XD}$ as a function of $k$. The largest deviation from unity reaches 0.2 at scales $\sim 2-5h\mathrm{Mpc}^{-1}$; the deviations at the same scales can be visually distinguished in Figure \ref{fig:simmodel}.

Beside the realization that has been used for training (realization A), we have two other with the identical physics, but different initial conditions (B and C). We apply the same analytical model to these two realizations and compare the power spectrum and cross correlation coefficients in Figure \ref{fig:simmodel}. The model performs equally well. Therefore, we can conclude that we did not over-trained the model. 

Also, in Figure \ref{fig:simmodel}, we show the power spectrum of the analytical model applied to a 160$\hMpc$ box, for which we do not have a numerical result. It illustrates how the analytical model can be expanded to the larger volumes.

Three stages captured in the Figure \ref{fig:simmodel} represent the beginning, the middle, and the very end of reionization. The first one is characterized by individual bubbles with only few galaxies within them. The model can reproduce the power spectrum; however, visually many bubbles are placed in the incorrect places. It happens because in this study we used only density fields. The model can be significantly improved if the halos would be cosidered as well. The second stage with overlapping bubbles correponds to the global ionization fraction reaching $\sim50\%$. At this stage bubbles contain multiple galaxies, and therefore the location of each individual halo becomes less important. We found this stage to be the most difficult to mimic (the deviations in correlation coefficient are the largest). Finally, the third presented stage shows the moment right after the reionization was complete. Only filaments remain partially neutral. At this stage the most uncertain are the small scales. It shows the limitations of our approach in which overdensities are directly linked with ionization fractions.

In \S\ref{subsec:discussion} we speculated that the shape of estimated barriers tells us that the most of the information lies in the range of scales 0.5-2 $h^{-1}$Mpc. However, here we showed that the largest deviations occur on the scales just above this range. We interpret it as the limitation of the barrier approach. Since the excursion set formalism is only an approximation to the numerical simulation, it is only capable of capturing the correlations partially.
%----------------------
%\subsection{Limits of analytical models}
%\label{subsec:limits_of_analytical_models}
%----------------------

%We performed all training forgetting about physical processes. Now we make a %step back and look what are the problematic regions in our analytical model %and look for reasons why it does not match the simulation.

%---------------
\section{Conclusions}
\label{sec:conclusions}
%---------------

In this paper we limit ourselves to the approach based on the excursion set formalism; however, having all of the available information from the phase space we could construct any mathematical model, for instance, based on the likelihood to the given median trajectories. The advantage of barrier approach is the similarity with existing and widely used analytical models. The excursion set formalism is not the only analytical method that can be fitted into numerical simulation. 

We have performed a detailed comparison between the numerical high resolution simulation and approximate analytical methods. We studied how the framework of the analytical model (excursion set formalism) can help to statistically describe ionization fronts in the numerical simulation. We have shown that it is possible and highlights some physical properties which are usually lost or cannot be easily seen in other statistics like power spectrum.

The statistics we developed is descriptive enough to be able to reproduce reionization history from density field. It allows us to build a model which combines the framework of excursion set formalism \cite{Furlanetto_2004} for describing large scale structure of ionization fronts and threshold approach \cite{miralda2000reionization} for small scale neutral patches -- filaments -- inside ionized bubbles. We perform training of the model into a given numerical simulation, using the power spectrum as our main criteria. The fitted model is capable to reproduce the power spectrum of other simulation realizations, which were not used during the fitting.

Ideally, a theoretical model should be able to describe all of the mentioned observations, as discussed in \S\ref{sec:discussion}. Our approach allows to build a single model, and compare it versus all available observations.

We propose the following work flow. First, one runs a simulation with any physics included. The box might be relatively small and number of realization can be low. Such a numerical simulation allows one to model the Thompson optical depth, Lyman-$\alpha$ forest and galaxy luminosity function. Then, the semi-analytical model is trained on this numerical simulation using the described method. The resulted model can generate as many mock catalogs as needed in boxes with sizes exceeding the size of the original simulation. These extended boxes are suitable for generating mock catalogs for 21cm experiments and to generate polarized CMB signal.

We made our code publicly available, further details are available on \url{https://bitbucket.org/kaurov/211mm}.

\acknowledgments
This work was supported by the NSF grant AST-1211190. AK is supported by the Fermilab Fellowship in Theoretical Physics.
Fermilab is operated by Fermi Research Alliance, LLC, under Contract
No.~DE-AC02-07CH11359 with the United States Department of Energy.

%\appendix
%\section{Excursion set framework in the numerical simulation}
%\label{app:barrierinnum}
%-------------------

%\section{B. Numerical spherical filter in Fourier space}
%\label{app:fourier_filter}
%-------------------

%Applying smoothing onto a 3D uniform grid through Fourier space is a usual technical operation. The benefit of this method is computational efficiency. In this paper we use spherical filter; however, any filter can be applied with this method. The work flow consists of (1) taking the Fourier transform of the original field, then (2) multiplying it on the Fourier transform of a filter, and (3) taking the inverse Fourier transform of the product. The resulting field is smoothed with the filter of choice. 

%The spherical filter has an anlytical counterpart in the Fourier space. Therefore it is a common practice to use this solution in the step (2). However, the analytical solution imposed on a grid in a naive way leads to inaccurate results. Furthermore, it often leads to a numerical effect known as 'ringing'.

%In order to avoid these problems, we do not use an analytical filter. Instead, we first construct a spherical filter on the exact same grid as we use in our analysis (it is not perfectly spherical, especially when filter size is small). Then, we take Fourier transform of the filter numerically. As a result, we constructed a filter in the Fourier space, which does not lead to any 'ringing' in the real space.

%\section{Barrier estimation}
%\label{app:barrierest}
%-------------------

%\bibliographystyle{apj}
\bibliography{bibs}

\end{document}